# Substrate-transferred GaAs/AlGaAs crystalline coatings for gravitational-wave detectors: A review of the state of the art

G. D. Cole,[1,a] S. Ballmer,[2] G. Billingsley,[3] S. B. Cataño-Lopez,[1] M. Fejer,[4]
P. Fritschel,[5] A. M. Gretarsson,[6] G. M. Harry,[7] D. Kedar,[8] T. Legero,[9]
C. Makarem,[3] S. D. Penn,[10] D. Reitze,[3] J. Steinlechner,[11,12] U. Sterr,[9]
S. Tanioka,[2] G.-W. Truong,[1] J. Ye,[8] J. Yu[9]

[1]*Thorlabs Crystalline Solutions, Santa Barbara, CA, 93101, USA*
[2]*Department of Physics, Syracuse University, Syracuse, NY, 13244, USA*
[3]*LIGO Laboratory, California Institute of Technology, Pasadena, CA, 91125, USA*
[4]*Stanford University, Palo Alto, CA, 94309, USA.*
[5]*LIGO Laboratory, Massachusetts Institute of Technology, Cambridge, MA, 02142, USA.*
[6]*Department of Physics, Embry-Riddle Aeronautical University, Prescott, AZ, 86301, USA*
[7]*Department of Physics, American University, Washington, DC, 20016, USA*
[8]*JILA, National Institute of Standards and Technology, University of Colorado Boulder, Boulder, CO, 80309, USA*
[9]*Physikalisch-Technische Bundesanstalt, 38116 Braunschweig, Germany*
[10]*Department of Physics, Hobart and William Smith Colleges, Geneva, NY, 14456, USA*
[11]*Maastricht University, 6200 MD Maastricht, The Netherlands*
[12]*Nikhef, 1098 XG Amsterdam, The Netherlands*

In this Perspective we summarize the status of technological development for large-area and low-noise substrate-transferred GaAs/AlGaAs (AlGaAs) crystalline coatings for interferometric gravitational-wave (GW) detectors. These topics were originally presented in a workshop[†] bringing together members of the GW community from the laser interferometer gravitational-wave observatory (LIGO), Virgo, and KAGRA collaborations, along with scientists from the precision optical metrology community, and industry partners with extensive expertise in the manufacturing of said coatings. AlGaAs-based crystalline coatings present the possibility of GW observatories having significantly greater range than current systems employing ion-beam sputtered mirrors. Given the low thermal noise of AlGaAs at room temperature, GW detectors could realize these significant sensitivity gains, while potentially avoiding cryogenic operation. However, the development of large-area AlGaAs coatings presents unique challenges. Herein, we describe recent research and development efforts relevant to crystalline coatings, covering characterization efforts on novel noise processes, as well as optical metrology on large-area (~10 cm diameter) mirrors. We further explore options to expand the maximum coating diameter to 20 cm and beyond, forging a path to produce low-noise AlGaAs mirrors amenable to future GW detector upgrades, while noting the unique requirements and prospective experimental testbeds for these novel materials.

---

[†]This Perspective serves as a summary of the AlGaAs Workshop, held at American University, Washington DC USA Aug. 15-17, 2022.
[a] Author to whom correspondence should be addressed. Electronic mail: gcole@thorlabs.com.

Thermal noise in high-reflectivity optical interference coatings is a limiting noise source in precision interferometric systems. The pioneering theoretical work on thermal noise by Callen and Greene [1] was introduced to the GW community by Saulson [2,3], as well as Braginsky and collaborators [4]. In 1998, Yuri Levin [5] identified coating thermal noise (CTN) as a potential limiting noise source for gravitational wave detectors. Harry, et al. [6] measured the elastic loss in the Initial LIGO coatings and confirmed that CTN would limit the sensitivity of Advanced LIGO [7]. A collaboration of Syracuse, Glasgow, Stanford, and the LIGO Lab determined the source of the loss to be the high-index material [8] and designed the coating [9] used in Advanced LIGO to make the first direct detection of these ripples in space-time [10].

For the past 20 years, a concerted research effort has sought to reduce CTN by identifying coating materials exhibiting both low levels of optical and elastic losses. The first decade of this work is summarized in Ref. 11. In the Advanced LIGO interferometers, coating Brownian noise limits the achievable strain sensitivity in the most sensitive frequency band around 100 Hz [12]. Similarly, this noise source impacts the stability of ultrastable optical resonators, placing a limit on the minimum linewidth achievable in lasers employed for cutting-edge optical atomic clocks [13,14]. This was initially explored in cavity-stabilized laser systems owing to theoretical work by Numata [15], followed by measurements on cm-length reference cavities by Notcutt and colleagues [16]. Exploratory efforts focusing on alternative materials with these same requirements were also carried out with micrometer-scale systems in the burgeoning field of cavity optomechanics [17]. Early work in this field ultimately led to the development of the substrate-transferred GaAs/AlGaAs (AlGaAs) crystalline coatings as described herein. A key motivation of these efforts is the potential for significant performance enhancements in GW detectors, owing to the low elastic losses and correspondingly low Brownian noise of these mirrors. As shown in Figure 1, in a model LIGO-based interferometer employing crystalline mirrors, the achievable strain sensitivity at 100 Hz is $1.1 \times 10^{-24} /\sqrt{\text{Hz}}$, representing a 3.6× improvement over the Advanced LIGO design target at the same frequency ($4.0 \times 10^{-24} /\sqrt{\text{Hz}}$) [12]. This results in a significant enhancement in the astrophysical reach for a binary neutron star merger, from 175 Mpc for the current design target to 600 Mpc for this proposed upgrade with AlGaAs-based crystalline coatings—yielding a factor of ~40 increase in detection rates.

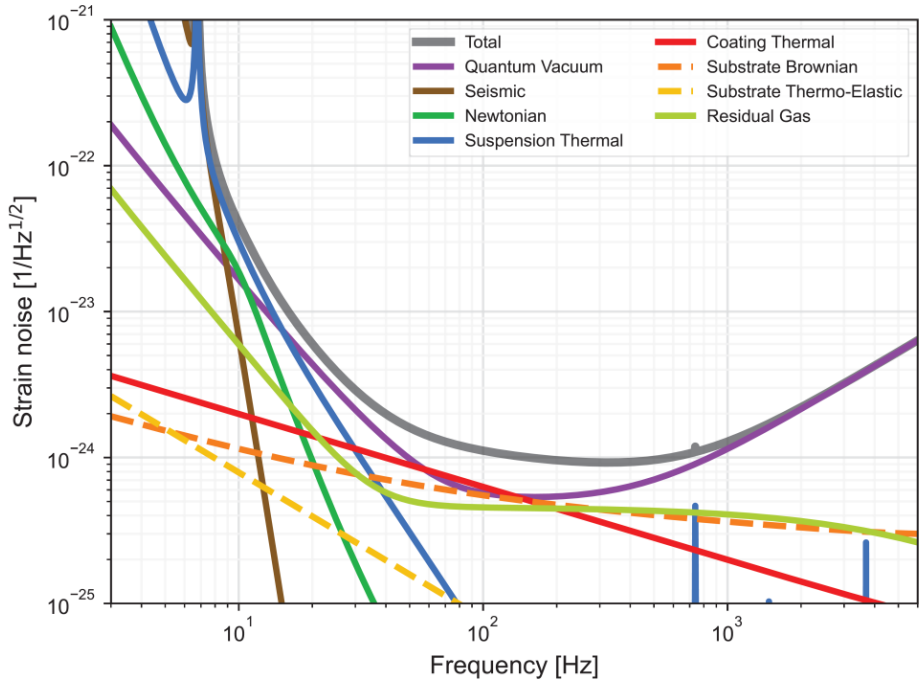

Fig. 1. Target strain sensitivity, as limited by fundamental noise sources, for a potential upgrade to the LIGO interferometers using AlGaAs crystalline coatings and operating at room temperature. The CTN curve (red) is calculated using a mechanical loss angle of $6.2\times10^{-6}$, based on direct thermal noise measurements at MIT[*1], and a beam radius of 5.5 cm on the end test masses and 4.5 cm on the input test masses; these beam sizes are compatible with a 30 cm diameter AlGaAs coating. The test masses are 100 kg fused silica (substrate noises), suspended with fused silica fibers (suspension thermal noise). The quantum vacuum noise derives from 1.5 MW of laser power in each arm cavity (1.06 µm wavelength), combined with 10 dB of effective vacuum squeezing at all frequencies, which is realized using a narrow linewidth filter cavity that appropriately rotates the squeezing angle as a function of frequency. The noise due to residual gas in the vacuum system arises from damping of the suspended test masses at frequencies below 50 Hz, and from scattering of the laser beams in the 4 km arms at frequencies above 50 Hz. The Newtonian noise is due to density perturbations in the earth close to the test masses, producing fluctuating gravitational forces. The seismic noise, in contrast, is the earth vibration that couples through the seismic isolation and test mass suspensions.

In this Perspective we provide a detailed account of the status of AlGaAs-based crystalline coatings as a solution to the coating Brownian noise problem. We begin with a brief historical overview of crystalline multilayers in cavity optomechanics experiments from ~2007-2012. We then discuss the transition of this technology to precision metrology applications with the development of centimeter-scale AlGaAs coatings for ultrastable optical reference cavities. Recent findings from key partners at national metrology labs point to novel noise processes in these coatings at cryogenic temperatures. Exploring size scaling,

---

[*1] Personal communication, S. Gras and N. Demos, MIT.

we cover preliminary results for crystalline mirrors at diameters up to 10 cm, with discussions relevant to expanding to 20 cm and beyond, covering the optical properties of these single-crystal films in terms of their absorption, scatter, birefringence, and surface uniformity. Given the opaque nature of AlGaAs coatings in the visible range, alternative lock acquisition schemes must be defined; one potential solution is presented here. Next is an overview of experimental testbeds that would enable detailed metrology of large-area crystalline mirrors. Finally, a brief overview of paths forward in terms of research and funding requirements is presented.

AlGaAs-based monocrystalline multilayers were first pursued as high-performance micromechanical resonators for cavity optomechanics experiments. This compound semiconductor material platform had historically been employed for microwave devices, as well as for micro-cavity-based optoelectronics devices. It was not until 2008 that measurements of the intrinsic elastic losses of AlGaAs were performed, revealing a unique combination of low optical and mechanical losses [18]. The realization of low elastic losses, represented by the mechanical loss angle, $\phi$ (or conversely, the mechanical quality factor, $Q = \frac{1}{\phi}$), was paramount to reaching the quantum regime in cavity optomechanics. Amorphous ion-beam sputtered (IBS) multilayers, while capable of high reflectivity, exhibited $Q$ values at the few thousand level (corresponding $\phi$ of a few ×10$^{-4}$). In comparison, crystalline Bragg mirrors, owing to their improved structural order, show a significant improvement, with measured $Q$ values in the range of ~ 20,000 to > 200,000 ($\phi$ at the low 10$^{-5}$ level or below) [18-21]. The low $Q$ values in IBS-deposited optical coatings presented a major roadblock in these efforts. This can be understood by the $Qf$ product, with $f$ being the mechanical eigenfrequency of the resonator. This eigenfrequency must exceed the thermal decoherence rate, yielding the condition: $Qf > k_B T_{bath}/h$ ($k_B$ – Boltzmann constant, $T_{bath}$ – system temperature, $h$ – Planck's constant), for the system to survive at least a single oscillation before a thermal phonon causes decoherence. This is a necessary requirement to prepare and detect nonclassical states of motion [22,23]. Ultimately, high-$Q$ AlGaAs-based micromechanical devices have been instrumental in studying fundamental aspects of quantum-limited interferometry [24,25].

The low-noise potential of these suspended micromirrors motivated the development of centimeter-scale reflectors based on substrate-transferred AlGaAs multilayers [26]. Production considerations for these novel coatings have been covered in detail elsewhere [27]. Given lattice matching constraints in epitaxial (crystal) growth, direct deposition is not possible, thus separate growth, microfabrication, and bonding is necessary to generate the coated optic. For low-loss macroscopic mirrors, optical quality is paramount. Each stage of the production process has the potential for defects, with the epitaxial growth stage contributing the largest share of defects. Since 2012, crystalline coatings, typically 5-20 mm in diameter, transferred to planar and curved bulk fused silica substrates have been realized. Other substrate materials have been successfully implemented including Si and $Al_2O_3$ (sapphire) for cryogenic reference cavities, as well as SiC, diamond, and YAG, for high-power laser systems. Optimized crystalline coatings with a radius of curvature as tight as 10 cm have demonstrated excess losses (scatter + absorption) below 2 ppm, with absorption as low as ~0.5 ppm observed between 1 μm and 1.5 μm. More recently, excess losses < 10 ppm have been demonstrated for mirrors operating near 4.5 μm [28,29]. The maturation of cm-scale crystalline coating production in the past decade has put this technology on par with IBS in terms of optical losses in the near-infrared spectral region, while exceeding the state-of-the-art in the mid-infrared (wavelengths from 2-5 μm). Standard mirrors are now commercially available, finding applications in cavity-stabilized lasers and in cavity-enhanced spectroscopy [30].

In time-and-frequency metrology, where high-finesse reference cavities employing crystalline coatings are becoming ubiquitous, the noise of AlGaAs multilayers has been closely studied and compared against theory. Room temperature cavity-stabilized lasers employing these coatings have been demonstrated to operate near the thermal noise floor [31,32], while turn-key systems capable of a fractional frequency instability $< 5 \times 10^{-16}$ are commercially available [33]. As the metrology community pushes optical oscillators to lower instabilities, the research focus has shifted to cryogenic systems. Progress on cryogenic reference cavities has matured to the point where the dominant noise contribution is CTN from the amorphous mirrors [34], making them ideal testbeds for probing low-temperature noise sources in AlGaAs coatings.

Two independent studies using silicon cavities with crystalline coatings at 1.5 μm have pioneered crystalline coating characterization at cryogenic temperatures. In both systems, the expected contributions from technical noise, and spacer and substrate thermal noise are well below the expected coating thermal noise (one cavity is 21 cm long and held at 124 K [35], the other is a 6 cm long cavity operated at 4 K and 16 K [36]). Interestingly, both systems have revealed hitherto unknown noise sources that can be manipulated by the polarization of the probe beam. Although it is well-known that coating birefringence leads to a static frequency splitting between orthogonal polarizations of the $TEM_{00}$ mode, the cryogenic testbeds additionally observe dynamic frequency fluctuations of the two polarization components that are anti-correlated. Additionally, the magnitude of this effect increases with the intracavity optical power. The resulting noise level is far above the coating thermal noise floor (20-40 dB depending on the cavity and the temperature) and the power spectral density acquires a slope steeper than $1/f$. Both birefringent modes of the cavity exhibit similar levels of frequency noise for equivalent optical conditions. However, if the two modes are addressed simultaneously, the anti-correlated frequency fluctuations can be averaged and suppressed [35,36], as in Figure 2. The residual noise after cancellation no longer scales with intracavity optical intensity, though it is still above the expected thermal noise level. This noise source is coherent between a $TEM_{10}$ and $TEM_{00}$ beam, implying a longer spatial correlation length than the spot size of ~1 mm. The source of this "global" noise is not yet understood and further investigations are ongoing. It also remains to be seen whether these measured noise scalings persist at higher frequency, as these cavities are optimized for the frequency range of 1 Hz and below, lower than the frequency band where coating thermal noise is relevant for Advanced LIGO, roughly 30-300 Hz.

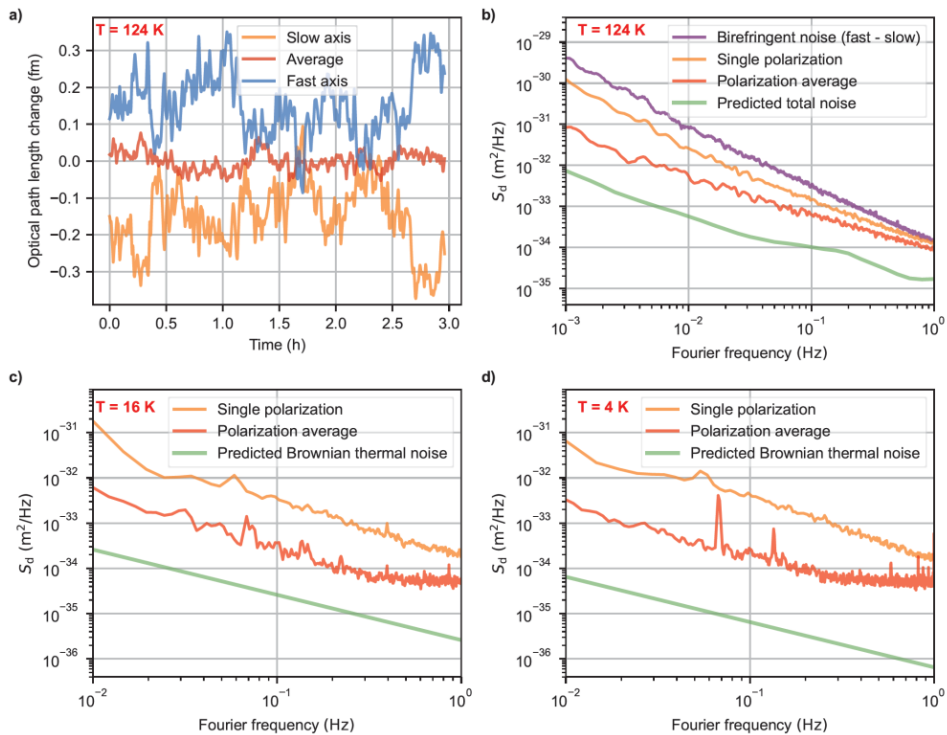

Fig. 2. Potential non-Brownian CTN observed in ultrastable cryogenic Si reference cavities with AlGaAs coatings. (a) Optical path length fluctuations of a 21-cm long Si cavity measured for the two polarization eigenmodes of the $TEM_{00}$ resonance (blue, orange). (b) Power spectral densities, $S_d$, of the length fluctuations of the cavity. The plot in this panel includes the birefringent noise (purple), the noise of an individual polarization eigenmode (orange), as well as the average of the two polarizations (red). The sum of technical noise and the measured Brownian thermal noise limit for the $TEM_{00}$ mode (green) is included for comparison. (c) and (d) Power spectral densities of the length fluctuations of a 6-cm long Si cavity at 16 K (c) and 4 K (d). Shown are the frequency stability of the individual polarization eigenmodes (orange), the average of two polarization eigenmodes (red), and the predicted Brownian thermal noise (green). As explained in the main text, while the average polarization can be used to suppress the birefringence fluctuations, the residual noise remains above the expected CTN level for these systems. Figure reproduced from Ref. 35.

It is important to note that the observed birefringence in crystalline coatings is an "extrinsic" effect, as 100-oriented GaAs is optically symmetric. The current conjecture is that non-uniform strain relaxation, upon cooling from the growth temperature drives the symmetry breaking [37]. The coatings are grown at elevated temperature (~600 ºC) leaving a compressive residual stress of ~100 MPa upon cooling. It is possible to tailor the strain by alloying the multilayer with In or P (e.g., GaAsP, InGaP, InGaAs, etc.) [38]. Careful measurement of the optomechanical properties of these materials would be necessary. The observed birefringence, measured from cavity mode splitting and corresponding to the accumulated difference in phase on reflection between the fast and slow polarizations, $\Delta\theta=\theta_f-\theta_s$, are all within a similar range, roughly $2\times10^{-3}$ radians, and are temperature and substrate independent. Furthermore, thermal cycling does not affect the mode splitting in cryogenic coatings.

Beyond studies of the static and dynamic (fluctuating) components of birefringence, there is a need to investigate potential thermally-induced effects. Owing to the anisotropic nature of AlGaAs coatings, radial thermal gradients will induce shear strains in the crystal. Assuming the mirror is a flat, half-infinite disk and the coating face is parallel to the [001] crystal plane, if we choose the $x$ and $y$ coordinate axes to lie in the [010] and [100] planes respectively, then the magnitude of the induced birefringence is largest along lines at +/- 45º to the coordinate axes. In addition, the orientation of the principal axes varies with azimuthal angle across the face of the mirror. Assuming a 100 µm diameter perfect absorber and a mirror irradiance at the level of Advanced LIGO, the effect may be similar in magnitude to the static birefringence seen in AlGaAs mirrors; several point absorbers could combine to impart a significant effect and must be investigated further.

Challenges with large-area mirror production are being tackled to extend the application of crystalline mirrors from advanced reference cavities to GW-detection. Studies on 2" and 3" (50.8-76.2 mm) diameter test mirrors have yielded: i) mean absorption < 1 ppm, ii) mean total integrated scatter (TIS): < 10 ppm, and iii) coating thickness variation (rms): < 100 ppm [39,40]. Ongoing efforts involve the production 10 cm and ultimately 20 cm diameter test mirrors, the latter representing the largest continuous crystalline coatings that can be produced, limited by the availability of base substrates for epitaxial growth. In terms of the observed optical properties in large coatings, there appears to be a greater number of scattering centers compared with the best IBS coatings, but the background total integrated scatter level away from larger (> 20 µm) scatterers is comparable. These scattering centers may be caused by "oval defects" arising from spitting of the gallium source during deposition, generating crystallites that locally disrupt the structure. It is not known whether these defects are absorbing; thus, distinguishing pure scattering centers from local absorbers is an important task. Both are a source of optical loss but have different impact on interferometer. Similarly, a bidirectional reflectance distribution function analysis of larger point defects will be useful for estimating the effect of rare but large scatterers versus frequent but small scatterers on interferometer scatter noise.

Recent results from Caltech include surface maps and scattering data from the first 10 cm diameter AlGaAs mirror (transferred to a 10-mm thick planar synthetic fused silica substrate), see Panel a of Figure 3 for more details. Surface figure measurements were made on the bare substrate before coating and on also the final coated mirror. After coating, the surface appears to have gained 4 nm of astigmatism over an 80 mm diameter aperture. The radius of the coated substrate changed by –784 m, with a corresponding sagitta change of 270 nm (convex) over an 80 mm diameter aperture. This could be due to non-uniformity of the coating or stress imparted on the 10 mm thick substrate. Panels b and c of Figure 3 show the results of the scattering measurements performed on this test structure. The mean TIS is somewhat higher than for the smaller samples mentioned earlier. This is due to an increased number of relatively strong scatterers indicated by red points (Fig. 3c). Excluding these large scattering centers, the mean total integrated scatter of this initial test mirror was approximately 2 ppm higher than equivalent measurements on an Advanced LIGO ETM, with a point scatterer density of 86 $cm^{-2}$. Additional 10 cm diameter test mirrors are currently in production to ascertain the repeatability in optical performance of these first large mirrors.

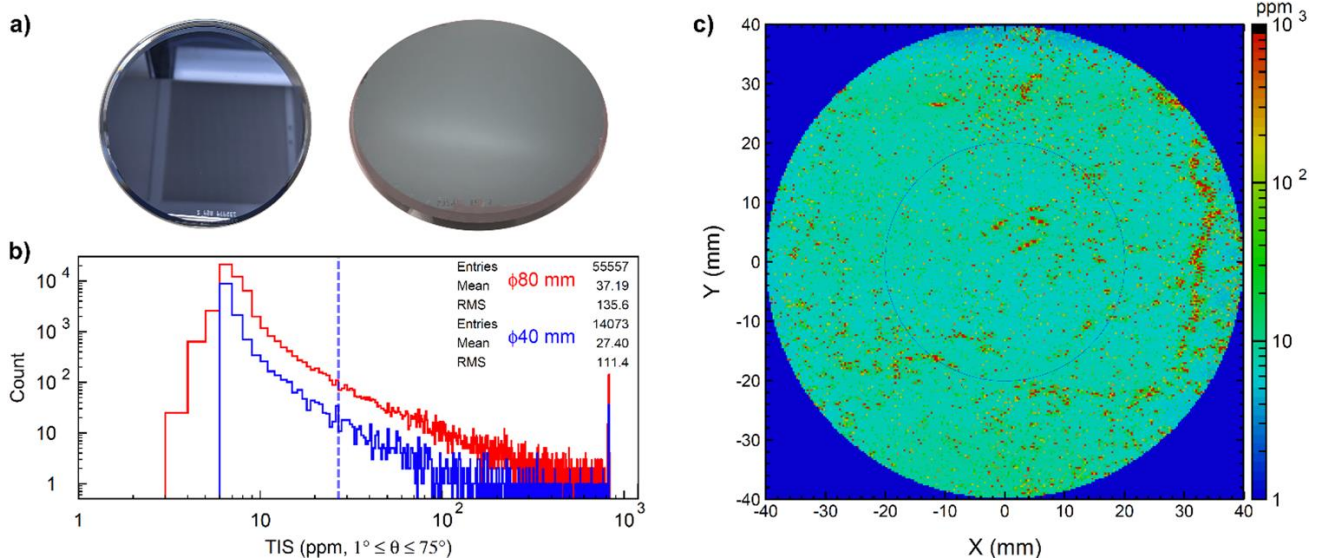

Fig. 3. Optical characterization of a first 10 cm diameter test mirror with substrate-transferred AlGaAs coatings. a) Photographs of the mirror backside (*left*), viewing the bond interface through the 10-mm thick fused silica substrate, and mirror frontside (*right*) following substrate and etch stop removal, leaving only the GaAs/AlGaAs multilayer on the fused silica substrate. b) Histogram of the TIS for apertures of 40 mm diameter (*blue*) and 80 mm diameter (*red*). The vertical dashed line shows the mean TIS for the 40 mm diameter aperture. The legend gives the means and standard deviations of both histograms. c) Scatter map obtained via an integrating sphere raster-scanned over the mirror surface. The thin blue circle shows the 40 mm diameter inner aperture referenced in panel b).

Given the lack of commercially available options, scaling to AlGaAs coating diameters beyond 20 cm will entail the growth of custom GaAs boules for waferization. Sticking with traditional wafer geometries, 30 cm diameter would be an obvious choice as a next step. In terms of multilayer epitaxy, production MBE systems have demonstrated sufficiently good optical performance and uniformity [27,40] and would not require customization. A dedicated system will be necessary to produce high-performance prototype and deliverable optics capable of meeting the strict optical specifications of GW-detectors [41]. At the bonding stage, commercial vendors have demonstrated the production of silicon-on-insulator wafers up to 45-cm diameter [42]. However, such systems are typically limited to a total bonded thickness of a few mm, while GW-relevant optics exhibit high mass (~40 kg) and much greater thickness (~20 cm), entailing modified tooling and unique challenges in production.

If sufficiently large mirrors with the desired performance metrics can be produced, it will then be necessary to explore impacts on the overall system operation. For instance, given the narrow bandgap of the high index GaAs layers, traditional cavity arm locking with frequency-doubled 532 nm light is no longer an option [43]; thus, alternative locking schemes must be developed. Given the long-wavelength transparency of AlGaAs, a dichroic coating with sufficient reflectance at both 1064 nm as well as an auxiliary wavelength of 2128 nm may be used. The auxiliary 2128 nm beam would be generated from 1064 nm beam, which is phase-locked to the main laser, using a degenerate optical parametric oscillator (DOPO) instead of a conventional frequency doubling [44]. If low-noise dichroic AlGaAs coatings can be produced, then a locking system can be realized with minimal risk. In principle such a coating design is possible, and based on recent optical loss measurements, at 4.5 μm, absorption at the 1 ppm level or less is expected for wavelengths near 2 μm [28,29].

Possible options for installing AlGaAs-coated input and end test masses (ITMs and ETMs) as an upgrade to the existing 4 km Advanced LIGO interferometers will require minimum coating diameters of 21-22 cm to exceed current requirements on coating thermal noise. Thus, the use of 20 cm GaAs wafers for both test masses is not possible without radical modifications and custom boules are needed. However, mixed mirror sizes could be employed to avoid this, with larger IBS-coated ITMs focusing a smaller spot onto AlGaAs coated ETMs. Such a design is limited by the size of the beamsplitter but has the advantage of keeping the power-recycling and signal-recycling cavity designs mostly unchanged (though it would require reshaping the anti-reflective side of the ITM to form a lens). Similar "mixed mirror" solutions leveraging 20 cm diameter AlGaAs will be evaluated for noise, alignment stability, resonances of higher order modes in the arms, sideband resonances in the arms, etc. These designs have the potential to serve as technology demonstrators for next-generation instruments.

As with cutting-edge ultrastable laser efforts, third-generation GW detectors such as the Einstein Telescope (ET) are proposing to incorporate cryogenics. To maintain compatibility with available growth substrates, the ET low frequency interferometer could potentially implement similar mixed mirror designs, including cooled ETMs with a 13 cm beam and 70 cm diameter IBS coatings and ITMs having a 4 cm spot size with 20 cm diameter AlGaAs coatings (with or without cryogenic cooling depending on the ultimate outcome of the cryogenic crystalline coating noise). With cryogenic cooling, this geometry could employ ultrapure float-zone silicon substrates for the ITMs, which are currently available up to 20 cm diameter.

Open questions relevant to AlGaAs in future GW detectors may go beyond that which can be answered in table-top experiments. These include: (a) an accurate wideband (frequency) measurement of coating noise; (b) successful production of larger than 'lab-scale' mirrors, spanning substrate procurement, polishing, and bonding, to integration with relevant suspension systems; and (c) investigations of large-area coating performance when integrated in a complex and sensitive system at high laser power. Several platforms will be available within the gravitational-wave community in the near-term for such efforts:

i) 10 m prototype at the AEI in Hannover, Germany [45], operating at 1064 nm and room temperature. A key aim is to investigate and overcome the standard quantum limit [25] and the use of low-noise AlGaAs-coated, 4.8 cm diameter mirrors have been proposed.

ii) Gingin prototype in Western Australia [46] will investigate high-power effects in silicon mirrors at a wavelength near 2 μm. This is a three-phase project: (1) 7 m Fabry Perot cavity at 5 W laser power and 3 mm beam diameter with interchangeable fused silica and silicon mirrors, (2) 72 m Fabry Perot cavity using silicon mirrors with 10 cm diameter AlGaAs coatings and ~1 cm beam diameter, (3) 23 kW laser power in the arm cavities at 123 K. This system can be used to explore the impact of point absorbers, wide angle scattering, thermal distortions and birefringence of coated and uncoated silicon substrates, and possibly electro-optic and non-linear effects at high laser power.

iii) There are currently two cryogenic protoypes under development, the ET-pathfinder in Maastricht, Netherlands [47] and a cryogenic upgrade to the 10 m prototype in Glasgow, UK [48], with the aim of testing technologies for low temperature operation of future GW detectors. Parameters such as 1.5 μm and 2 μm laser wavelengths at temperatures of ~120 K and ~15-20 K are planned, using Si substrates for the mirrors. These systems could be ideal platforms to test large-area AlGaAs coatings beyond ongoing efforts with cm-scale reference cavities.

These platforms will be instrumental in confirming the viability of AlGaAs coatings in these unique astronomical instruments.

We have outlined the historical background in the initial development of, as well as the current status and potential paths forward for, AlGaAs-based crystalline coatings. These unique coatings exhibit promising optomechanical properties for enhanced sensitivity in GW detection and thus demand further investigation. We end by clarifying that the cost and timeline to realize the large-diameter production capabilities above (custom base wafers, epitaxy, and bonding) is comparable to the development of other important subsystems such as seismic isolation [49] and quantum squeezing [50], which is an appropriate comparison in that coating thermal noise is the limiting noise in the most sensitive frequency band of second-generation GW detectors [7]. This can also be compared to potential budgetary savings realized by putting off or even eliminating the need to develop cryogenics for future detectors [51]. A proposed timeline (available to LIGO, Virgo, and KAGRA members) has been developed that would allow GW-relevant AlGaAs coatings to be realized within a span of 5 years. This will allow for AlGaAs to be considered for upgrades to the Advanced LIGO detectors.

## ACKNOWLEDGEMENTS

This research was supported by the National Science Foundation through the following grant awards: American University (PHYS-2012017, PHYS-2011787), Caltech (PHY-1764464), HWS (PHY-1912699, PHY-2011688, PHY-2208079), M.I.T. (PHY-1764464), Embry-Riddle Aeronautical University (PHY- 2110598), Syracuse University (PHY-2011723, PHY-2207640). J. Steinlechner acknowledges the support of ETpathfinder (Interreg Vlaanderen-Nederland), E-TEST (Interreg Euregio Meuse-Rhine), the Province of Limburg, and support by the NWO Talent Programme Vidi 2020, project number VI.Vidi.203.062. D. Kedar and J. Ye acknowledge support from NSF PHY-1734006, NIST, and AFRL. T. Legero, U. Sterr and J. Yu acknowledge support by the Project 20FUN08 NEXTLASERS, which has received funding from the EMPIR programme co-financed by the Participating States and from the European Union's Horizon 2020 Research and Innovation Programme. A portion of this work was performed in the UCSB Nanofabrication Facility, an open access laboratory.

## DATA AVAILABILITY

The data that support the findings of this study are available from the corresponding author upon reasonable request.

# REFERENCES

[1] H. B. Callen and R. F. Greene, "On a Theorem of Irreversible Thermodynamics," Phys. Rev. **86**, 702 (1952).
[2] P. R. Saulson, "Thermal noise in mechanical experiments", Phys. Rev. D **42**, 2437 (1990).
[3] G. I. González and P. R. Saulson, "Brownian motion of a mass suspended by an anelastic wire," J. Acoust. Soc. Am. **96**, 207 (1994)
[4] V. B. Braginsky, V. P. Mitrofanov, and V. I. Panov, *Systems with Small Dissipation.* University of Chicago Press 1985.
[5] Y. Levin, "Internal thermal noise in the LIGO test mass: a direct approach," Phys. Rev. D **57**, 649 (1998).
[6] G. M. Harry, A. M. Gretarsson, P. R. Saulson, S. E. Kittelberger, S. D. Penn, W. J. Startin, S. Rowan, M. M. Fejer, D. R. M. Crooks, G. Cagnoli, J. Hough, and N. Nakagawa, "Thermal noise in interferometric gravitational wave detectors due to dielectric optical coatings," Classical Quant. Grav. **19**, 897 (2002)
[7] B. P. Abbott et al. (The LIGO Scientific Collaboration), "GW150914: The Advanced LIGO Detectors in the Era of First Discoveries," Phys. Rev. Lett. **116**, 131103 (2016).
[8] S. D. Penn, P. Sneddon, H. Armandula, J. C. Betzwieser, G. Cagnoli, J. Camp, D. R. M. Crooks, M. Fejer, A. M. Gretarsson, G. M. Harry, J. Hough, S. E. Kittleberger, M. J. Mortonson, R. Route, S. Rowan, P. R. Saulson, and C. C. Vassiliou, "Mechanical Loss in Tantala/Silica Dielectric Optical Coatings," Classical Quant. Grav. **20**, 2917 (2003).
[9] G. M. Harry et al., "Titania-doped tantala/silica coatings for gravitational-wave detection," Classical Quant. Grav. **24**, 405 (2007).
[10] B. P. Abbott et al. (The LIGO Scientific Collaboration and Virgo Collaboration), "Observation of Gravitational Waves from a Binary Black Hole Merger," Phys. Rev. Lett. **116** 061102 (2016).
[11] Optical coatings and thermal noise in precision measurement, ed. G. Harry, T. P. Bodiya, and R. DeSalvo, Cambridge University Press 2012.
[12] Aasi et al (The LIGO Scientific Collaboration), "Advanced LIGO," Class. Quantum Grav. **32**, 074001 (2015).
[13] D. G. Matei, T. Legero, S. Häfner, C. Grebing, R. Weyrich, W. Zhang, L. Sonderhouse, J. M. Robinson, J. Ye, F. Riehle, and U. Sterr, "1.5 μm lasers with sub-10 mHz linewidth," Phys. Rev. Lett. **118**, 263202 (2017).
[14] E. Oelker, R. B. Hutson, C. J. Kennedy, L. Sonderhouse, T. Bothwell, A. Goban, D. Kedar, C. Sanner, J. M. Robinson, G. E. Marti, D. G. Matei, T. Legero, M. Giunta, R. Holzwarth, F. Riehle, U. Sterr, and J. Ye, "Demonstration of $4.8 \times 10^{-17}$ stability at 1 s for two independent optical clocks," Nat. Photonics **13**, 714 (2019).
[15] K. Numata, A. Kemery, and J. Camp, "Thermal-noise limit in the frequency stabilization of lasers with rigid cavities", Phys. Rev. Lett. **95**, 173602 (2004).
[16] L.-S. Ma, A. D. Ludlow, S. M. Foreman, J. Ye, and J. L. Hall, "Contribution of thermal noise to frequency stability of rigid optical cavity via Hertz-linewidth lasers," Phys. Rev. A **73**, 031804 (2006).
[17] M. Aspelmeyer, T. J. Kippenberg, and F. Marquardt, "Cavity optomechanics," Rev. Mod. Phys. **86**, 1391 (2014).
[18] G. D. Cole, S. Gröblacher, K. Gugler, S. Gigan, and M. Aspelmeyer, "Monocrystalline $Al_xGa_{1-x}As$ heterostructures for high-reflectivity high-Q micromechanical resonators in the megahertz regime," Appl. Phys. Lett. **92**, 261108 (2008).
[19] G. D. Cole, I. Wilson-Rae, M. R. Vanner, S. Gröblacher, J. Pohl, M. Zorn. M. Weyers, A. Peters, and M. Aspelmeyer, "Megahertz monocrystalline optomechnaical resonators with minimal dissipation," 23[rd] IEEE International Conference on MEMS, Hong Kong, China, 24–28 January 2010, TP133, pp. 847-850.
[20] G. D. Cole, Y. Bai, M. Aspelmeyer, E. A. Fitzgerald, "Free-standing $Al_xGa_{1-x}As$ heterostructures by gas-phase etching of germanium," Appl. Phys. Lett. **96**, 261102 (2010).
[21] G. D. Cole, "Cavity optomechanics with low-noise crystalline mirrors," SPIE Optics & Photonics, Optical Trapping and Optical Micromanipulation IX, San Diego, CA, USA, 12–16 August 2012 [8458-07].
[22] W. Marshall, C. Simon, R. Penrose, and D Bouwmeester, "Towards Quantum Superpositions of a Mirror," Phys. Rev. Lett. **91**, 130401 (2003).
[23] Y. Chen, "Macroscopic quantum mechanics: theory and experimental concepts of optomechanics," J. Phys. B: At. Mol. Opt. Phys. **46** 104001 (2013).
[24] J. Cripe, N. Aggarwal, R. Lanza, A. Libson, R. Singh, P. Heu, D. Follman, G. D. Cole, N. Mavalvala, and T. Corbitt, "Measurement of quantum back action in the audio band at room temperature," Nature **568**, 364 (2019).
[25] T. Cullen, R. Pagano, J. Cripe, S. Sharifi, M. Lollie, S. Aronson, H. Cain, P. Heu, D. Follman, N. Aggarwal, G. D. Cole, and T. Corbitt, "Surpassing the standard quantum limit using an optical spring," preprint arXiv:2210.12222.
[26] G. D. Cole, W. Zhang, M. J. Martin, J. Ye, and M. Aspelmeyer, "Tenfold reduction of Brownian noise in high-reflectivity optical coatings," Nature Photonics **7** 644 (2013).

[27] G. D. Cole, W. Zhang, B. J. Bjork, D. Follman, P. Heu, C. Deutsch, L. Sonderhouse, J. Robinson, C. Franz, A. Alexandrovski, M. Notcutt, O. H. Heckl, J. Ye, and M. Aspelmeyer, "High-performance near- and mid-infrared crystalline coatings," Optica **3**, 647 (2016).
[28] G. Winkler, L. W. Perner, G.-W. Truong, G. Zhao, D. Bachmann, A. S. Mayer, J. Fellinger, D. Follman, P. Heu, C. Deutsch, D. M. Bailey, H. Peelaers, S. Puchegger, A. J. Fleisher, G. D. Cole, and O. H. Heckl, "Mid-infrared interference coatings with excess optical loss below 10 ppm," Optica **8**, 686 (2021).
[29] G.-W. Truong, L. W. Perner, G. Winkler, S. B, Cataño-Lopez, C. Nguyen, D. Follman, O. H. Heckl, and G. D. Cole, "Transmission-dominated mid-infrared supermirrors with finesse exceeding 200 000," preprint arXiv:2209.09902.
[30] B. J. Bjork, T. Q. Bui, O. H. Heckl, P. B. Changala, B. Spaun, P. Heu, D. Follman, C. Deutsch, G. D. Cole, M. Aspelmeyer, M. Okumura, and J. Ye, "Direct frequency comb measurement of OD + CO → DOCO kinetics," Science **354**, 444 (2016).
[31] S. Herbers, S. Häfner, S. Dörscher, T. Lücke, U. Sterr, and C. Lisdat, "Transportable clock laser system with an instability of $1.6 \times 10^{-16}$," Opt. Lett. **47**, 5441 (2022).
[32] M. Kelleher, F. Quinlan, manuscript in preparation
[33] M. Brekenfeld, B. Rauf, S. Saint-Jalm, G. D. Cole, G.-W. Truong, M. Lessing, A. Fricke, M. Fischer, M. Giunta, R. Holzwarth, "Rack-mounted ultrastable laser system for Sr lattice clock operation," Conference on Lasers and Electro-Optics (CLEO), San Jose, CA, USA, 15–20 May 2022.
[34] J. M. Robinson, E. Oelker, W. R. Milner, W. Zhang, T. Legero, D. G. Matei, F. Riehle, U. Sterr, and J. Ye, "Crystalline optical cavity at 4 K with thermal noise limited instability and ultralow drift," Optica **6**, 240 (2019).
[35] J. Yu, D. Kedar, S. Häfner, T. Legero, F. Riehle, S. Herbers, D. Nicolodi, C. Y. Ma, J. M. Robinson, E. Oelker, J. Ye, and U. Sterr, "Excess noise in highly reflective crystalline mirror coatings," preprint arXiv: 2210.15671.
[36] D. Kedar, J. Yu, E. Oelker, A. Staron, W. R. Milner, J. M. Robinson, T. Legero, F. Riehle, U. Sterr, and J. Ye, "Frequency stability of cryogenic silicon cavities with semiconductor crystalline coatings," preprint arXiv: 2210.14881
[37] M. Bückle, V. C. Hauber, G. D. Cole, C. Gärtner, U. Zeimer, J. Grenzer, and E. M. Weig, "Stress control of tensile-strained In$_{1-x}$Ga$_x$P nanomechanical string resonators," Appl. Phys. Lett. **113**, 201903 (2018).
[38] G. D. Cole, P.-L. Yu, C. Gärtner, K. Siquans, R. Moghadas Nia, J. Schmöle, J. Hoelscher-Obermaier, T. P. Purdy, W. Wieczorek, C. A. Regal, and M. Aspelmeyer, "Tensile strained In$_x$Ga$_{1-x}$P membranes for cavity optomechanics," Appl. Phys. Lett. **104**, 201908 (2014).
[39] M. Marchiò, R. Flaminio, L. Pinard, D. Forest, C. Deutsch, P. Heu, D. Follman, and G. D. Cole, "Optical performance of large-area crystalline coatings," Opt. Express **26**, 6114 (2018).
[40] P. Koch, G. D. Cole, C. Deutsch, D. Follman, P. Heu, M. Kinley-Hanlon, R. Kirchhoff, S. Leavey, J. Lehmann, P. Oppermann, A. K. Rai, Z. Tornasi, J. Wöhler, D. S. Wu, T. Zederbauer, and H. Lück, "Thickness uniformity measurements and damage threshold tests of large-area GaAs/AlGaAs crystalline coatings for precision interferometry," Opt. Express **27**, 36731 (2019).
[41] G. Harry and G. Billingsley, "Fused Silica, Optics, and Coatings", in Advanced Gravitational-wave Detectors Volume 2, ed. D. Reitze, P. Saulson, and H. Grote, World Scientific (2019).
[42] D. McGrath, "EV Group to ship 450-mm wafer bonding tool," EE Times (2011).
[43] A. Staley et al. "Achieving Resonance in the Advanced LIGO Gravitational-Wave Interferometer." Classical Quant. Grav. **31**, 245010 (2014).
[44] C. Darsow-Fromm, M. Schröder, J. Gurs, R. Schnabel, and S. Steinlechner, "Highly efficient generation of coherent light at 2128 nm via degenerate optical-parametric oscillation," Opt. Lett. **45**, 6194 (2020).
[45] C. Gräf, S. Hild, H. Lück, B. Willke, K. A. Strain, S. Goßler, and K. Danzmann, "Optical layout for a 10 m Fabry–Perot Michelson interferometer with tunable stability," Class. Quantum Grav. **29** 075003 (2012).
[46] C. Zhao et al. "Gingin High Optical Power Test Facility," J. Phys.: Conf. Ser. **32**, 368 (2006).
[47] A Utina et al., "ETpathfinder: a cryogenic testbed for interferometric gravitational-wave detectors," Class. Quantum Grav. **39**, 215008 (2022).
[48] K. D. Skeldon, D. A. Clubley, B. W. Barr, M. M. Casey, J. Hough, S. D. Killbourn, P.W. McNamara, G. P. Newton, M. V. Plissi, D. I. Robertson, N. A. Robertson, K. A. Strain, and H. Ward, "Performance of the Glasgow 10 m prototype gravitational wave detector operating at λ=1064 nm". Phys. Lett. A **273**, 277 (2000).
[49] F. Matichard et al. "Advanced LIGO two-stage twelve-axis vibration isolation and positioning platform. Part 1: Design and production overview," Precis. Eng. **40**, 273 (2015).
[50] E. Oelker, G. Mansell, M. Tse, J. Miller, F. Matichard, L. Barsotti, P. Fritschel, D. McClelland, M. Evans, and N. Mavalvala, "Ultra-low phase noise squeezed vacuum source for gravitational wave detectors", Optica **3**, 682 (2016).
[51] S. Hild et al, (The Einstein Telescope Collaboration), "Sensitivity studies for third-generation gravitational wave observatories," Classical Quant. Grav. **28**, 094013 (2011)